\documentclass[10pt]{article}
\usepackage{authblk}

\usepackage{amssymb}
\usepackage{graphicx}
\usepackage{hyperref}
\usepackage{algorithm,float}
\usepackage[noend]{algpseudocode}
\usepackage[numbers]{natbib}
\usepackage{booktabs, makecell, longtable}
\usepackage{textcomp}
\usepackage[font=small,skip=0pt]{caption}
\makeatletter
\def\BState{\State\hskip-\ALG@thistlm}
\usepackage[margin=.5 in]{geometry}

\usepackage{booktabs}
\usepackage{multirow}
\usepackage[table,xcdraw]{xcolor}

\usepackage{amsmath}
\usepackage{pdfpages}
\usepackage{url}
\urldef{\mailsa}\path|{alfred.hofmann, ursula.barth, ingrid.haas, frank.holzwarth,|
	\urldef{\mailsb}\path|anna.kramer, leonie.kunz, christine.reiss, nicole.sator,|
	\urldef{\mailsc}\path|erika.siebert-cole, peter.strasser, lncs}@springer.com|

\begin{document}	
\UseRawInputEncoding 
\raggedbottom

	

	\title{A Deep Learning-Based Unified Framework for Red Lesions Detection on Retinal Fundus Images}

	\author[1]{Norah Asiri\thanks{norah.m.asiri@outlook.com}}
		\affil[1]{Computer and Information Science College, King Saud University, Riyadh, Saudi Arabia}
	 
	\author[2]{Fadwa Al Adel \thanks{ffaladel@pnu.edu.sa}}
	\affil[2]{Department of Ophthalmology, College of Medicine, Princess Nourah bint Abdulrahman University, Riyadh, Saudi Arabia}

	\maketitle

		\begin{abstract}


Red-lesions, i.e., microaneurysms (MAs) and hemorrhages (HMs), are the early signs of diabetic retinopathy (DR). The automatic detection of MAs and HMs on retinal fundus images is a challenging task. Most of the existing methods detect either only MAs or only HMs because of the difference in their texture, sizes, and morphology. Though some methods detect both MAs and HMs, they suffer from the curse of dimensionality of shape and colors features and fail to detect all shape variations of HMs such as flame-shaped HM. Leveraging the progress in deep learning, we proposed a two-stream red lesions detection system dealing simultaneously with small and large red lesions. For this system, we introduced a new ROIs candidates generation method for large red lesions on fundus images; it is based on blood vessel segmentation and morphological operations, and reduces the computational complexity, and enhances the detection accuracy by generating a small number of potential candidates. For detection, we proposed a framework with two streams. We used pre-trained VGGNet as a backbone model and carried out several extensive experiments to tune it for vessels segmentation and candidates generation, and finally learning the appropriate mapping, which yields better detection of the red lesions comparing with the state-of-the-art methods. The experimental results validated the effectiveness of the system in the detection of both MAs and HMs; it yields higher performance for per lesion detection; its sensitivity equals 0.8589 and good FROC score under 8 FPIs on DiaretDB1-MA reports FROC = 0.7518, and with SN = 0.7552 and good FROC score under 2,4 and 8 FPIs on DiaretDB1-HM, and SN = 0.8157 on e-ophtha with overall FROC = 0.4537 and on ROCh dataset with FROC = 0.3461 which is higher than the state-of-the art methods. For DR screening, the system performs well with good AUC on DiaretDB1-MA, DiaretDB1-HM, and e-ophtha datasets.

	\end{abstract}
 

	\section{Introduction}
Diabetic retinopathy (DR) is a chronic and constitutes the first cause of blindness in the working-age population. It is emerging as one of the most dreaded sight complications. The fundamental problem of DR is that it usually symptoms in its late phase becomes incurable, therefore the importance of early diagnosis procedures has arisen. However, this involves a remarkable difficulty in the health care due to high number of potential patients. Additionally, for an effective follow-up of specialists, an enormous availability of ophthalmologists needed connected to all cases and conditions.

DR can be classified into two main classes based on its severity: non-proliferative DR (NPDR) and proliferative DR (PDR) 	\cite{early1991grading,acharya2009computer}. The clinical feature of NPDR stage is at least one MA or HM with or without hard exudates. MAs are small bulges appeared in the blood vessels which may leak blood on retinal layers causing HM while DR progresses. This is very common in people with diabetes \cite{dr}.
		
Digital retinal imaging uses high-resolution systems to capture images of eye. This helps clinicians to determine the validity of retina and, at the same time, recognize and control eye diseases such as glaucoma, diabetic retinopathy and macular degeneration. In addition to early disease detection, these images provide a constant record of changes in retina. For example, such images can track the most subtle retinal changes and will help doctors and inform them about patient health. It is necessary to recognize retinal anomalies as soon as possible to prevent the development of potentially serious illnesses or even loss of vision. However, this involves a remarkable difficulty in the health care system due to many potential patients and a small number of experienced ophthalmologists. It motivated the need to develop automated diagnosis systems to assist in early diagnosis of DR.
	
It is important to develop an automatic system to assist in the pre-diagnosis of DR in diabetic patients to rapidly assess the retina and indicate if there are any lesions that must be treated. The outstanding performance of deep learning in various computer vision tasks motivated its application for medical image analysis, in particular, retinal fundus image analysis and as soon as a suitable amount of data is available. It has been applied to a variety of tasks, including diagnosis, detection, segmentation, controlling, monitoring and visualization of pathologies in retinal fundus images.
	
Deep learning, in particular,  convolutional neuronal networks (CNN), has become an increasingly important subject in artificial intelligence and has helped to make progress in areas as diverse as object recognition. Employing CNN for DR diagnosis needs a huge amount of data to overcome the overfitting problem and ensure proper convergence \cite{tajbakhsh2016convolutional}. The expert annotations of data are expensive, and the appearance of lesions is not the default case. One advantage of CNNs is the ability to transfer the information embedded in the pre-trained CNNs. Transfer learning can speed up the learning process and enhance generalization \cite{erhan2010does}.

	
In this paper, we develop an automatic red lesions detection system for DR computer-aided diagnosis. It detects both microaneurysms (MA) and hemorrhages (HM) based on a deep learning approach. The proposed method deals with red lesions detection as an object detection problem; it finds the deep features of automatically generated small and large candidates to classify them into red lesions or non-red lesions. To detect small and large red lesions using the same system is a challenging task. We employ a two-stream approach for this issue. To extract potential candidates for small and large red lesions is a fundamental and difficult task in this approach. Because of the morphology, sizes, and textures of small and large red lesions, the same method does help to generate potential candidates. Because of this, we break up this task into two sub-tasks: candidates generation for small red lesions and large red lesions, and introduce a novel technique based on deep learning to generate large red lesion candidates.

The proposed method takes a fundus image as input, preprocesses it using contrast equalization (CE). Then it generates small red lesions candidates on the whole fundus image in an unsupervised manner using morphological operations. Afterward, it splits the enhanced image and the one with small red lesions candidates into patches of the same size for computational efficiency. Next, it extracts large red lesions candidates from patches by removing blood vessels using a fully convolutional network (FCN-8) and retrieves large potential lesions using a threshold mask. The patches with small and large red lesion candidates are fed into two subnets to learn discriminative features and detect the red lesions’ unique features. We validated the method on benchmark public datasets such as e-ophtha \cite{optha}, DiaretDB1 \cite{kalviainen2007diaretdb1}, ROCh\cite{rocds}, and a private dataset collected from the diabetes center of King Abdulaziz university hospital in Saudi Arabia.

\section{Related works}
	
Many methods have been proposed for the automatic detection of red lesions, i.e., MAs and HMs on fundus images. These methods can be broadly categorized into two classes depending on whether they are based on hand-engineered features or deep learning. The main framework followed by most of these methods consists of preprocessing, extraction of region proposals, their classification, and the refinement of their bounding boxes to detect and locate red lesions.

In red lesions detection, extraction of region proposals (i.e., candidates generation) plays an important role. The candidates' generation can be classified into categories: (i) brute force methods where any region is unconditionally considered as a candidate ((\citet{eftekhari2019microaneurysm,chudzik2018microaneurysm1}) and (ii) the methods which generate a small number of candidates (\citet{orlando2018ensemble,romero2019entropy,seoud2016red,long2020microaneurysms}). The main disadvantage of first-type methods is that they produce a large number of proposals, most of which are redundant and not related to red lesions and make training expensive in space and time. On the other hand, second type methods are more intelligent and faster, such as the method based on morphological operations proposed by \citet{orlando2018ensemble,seoud2016red,long2020microaneurysms}, region growth \citet{wu2017automatic,adal2014automated} and methods based on superpixel introduced by \citet{romero2019entropy}.

	\paragraph{Methods Based on Hand-Engineered Features}
 \citet{wu2017automatic} focus on MAs detection using the four-stages method. First, preprocessing step is applied to the green channel, which includes illumination equalization enhancement and smoothing. Afterward, MAs candidates are generated using peak detection and region growing. Then, local features such as Hessian matrix-based features, shape and intensity features, and other profile features are extracted. Finally, K-nearest neighbor (KNN) is used as a classifier. The overall FROC score reaches 0.273 on the e-ophtha MA dataset, which is the lowest compared to other works on the same dataset. Though this method is simple and does not include complex segments, KNN is sensitive to data scalability and irrelevant features.
	
 \citet{long2020microaneurysms} also focus on MAs detection. First, shade correction preprocessing is performed by applying the median filter on the green channel taking filter size larger than the maximal blood vessel width in the fundus image. Then, the resulting image is subtracted from the green channel, and the mean of the green channel is added to enhance contrast. Next, blood vessels are segmented using eigenvalues of the Hessian matrix. Afterward, MAs candidates are extracted using connected component analysis and shape characteristics. Then, directional local contrast (DLC) features are extracted from each candidate patch, and finally, Naive Bayes is used as a classifier. This method was assessed on e-ophtha and DiaretDB1-MA datasets, and the reported sensitivity value at the average 8 FPIs is 0.7 with an e-ophtha FROC score of 0.374 and DiaretDB1MA FROC score of 0.210. The main disadvantage of this method is the high dimensional DLC features, which lead to poor performance.
		 	 
		\citet{adal2014automated} introduced a three-stage method to detect MAs and dot HMs (small red lesions). First, a singular value decomposition-based contrast enhancement is used to reduce the shading effect while increasing the contrast of fundus images. Then, MAs candidates are extracted using descriptors of scale-invariant regions to detect the blob regions. Finally, a semisupervised learning strategy is used for classification. The method was trained using only a few manually labeled retinal images. The overall performance on DiaretDB1-MA, reported in terms of FROC score, is 0.184, which is very low.
		   
		  \citet{romero2019entropy} assume every dark region as a candidate. First, bright border artifact is removed by simulating wider aperture, illumination and color equalization, denoising, and contrast enhancement is applied as preprocessing. Then, pixels,  similar in color and texture,  are grouped in superpixels using the entropy rate superpixel method to separate different parts of the retina. The similarity is measured using 39 hand-crafted features to identify red lesions. After that, they use a three layers perceptron for classification. They used the 61 testing images of DiaretDB1 with 84.04\% sensitivity, 85\% specificity, and 84.45\% accuracy.

\citet{seoud2016red} proposed a method that differentiates between red lesions and blood vessels without segmenting blood vessels. First, illumination equalization, denoising, adaptive contrast equalization, and color normalization are applied as preprocessing. Then, dynamic shape features are used to define candidates after removing the optic disk. After that, random forest (RF) is used for classification. Though this method aims to detect both MAs and HMs, it fails to detect flame-shape HMs because of similarity with blood vessels. The method was validated on six datasets, and the overall FROC score on DiaretDB1 is 0.3540.

\citet{zhang2010detection} proposed a MAs detection method based on dynamic thresholding and multi-scale correlation filtering of Gaussian templates and 31 hand features such as intensity, shape, and response of a Gaussian filter on the green channel. First, MAs candidates are generated using the coarse level of the Gaussian template. Next, MAs are classified using the fine level of the Gaussian template. This method was evaluated on the ROCh training dataset and reported FROC score equal to 0.246. \citet{javidi2017vessel} proposed a two-stages MAs detection approach. First, they segment blood vessels using discriminative dictionary learning and sparse representation. After that, MAs candidates are generated using a 2D Morlet wavelet. Next, similar to vessel segmentation, a discriminative dictionary learning approach distinguishes MAs from non-MAs objects. This method has been tested on the ROCh training dataset and yielded an overall FROC score of 0.261.

 	\paragraph{Methods Based on Convolutional Neuronal Networks (CNN)}
	\citet{orlando2018ensemble} fused CNN features learning and hand-engineered features (HEF) and then used random forest (RF) to identify the lesion candidates. First, they used contrast equalization as preprocessing step. Then, candidates were extracted by isolating red lesions based on their properties, such as shape and number of pixels using morphological operations. They also used vessels segmentation to enhance candidates' extraction. They evaluated on  DiaretDB1 and e-ophtha datasets for per lesion evaluation and obtained FROC score of  0.3683 for e-ophtha, 0.3301 for DiaretDB1-MA and 0.5044 for DiaretDB1-HM. Although this method performs well with small red lesions, it fails to detect medium to large red lesions. Also, this method is time-consuming since it classifies each candidate box separately.
	
	\citet{eftekhari2019microaneurysm} also proposed a two-stage method for MAs detection. First, color normalization and retina background elimination are applied as preprocessing, then 101× 101 patches are extracted, and a CNN model is used to generate a probability map. In the second stage, another CNN model is used to classify each pixel as MA or non-MA. The sensitivity value at an average of 6 FPIs reaches 0.8 on the e-ophtha MA dataset with an FROC score of 0.471. Though this method has good performance, it is time-consuming since it has pixel-based classification.
	
	\citet{chudzik2018microaneurysm1} employed a fully convolutional network (FCN) to detect MAs in three stages. First, in preprocessing, the green channel is obtained and cropped around FOV, then the non-uniform illumination is normalized. After that, patches are extracted and divided into MA patches containing at least one MA and non-MA patches. Then, the FCN model is used for pixel-wise classification. This method was evaluated on the e-ophtha, DiaretDB1, and ROCh training datasets; it achieved FROC scores of 0.562, 0.392, and 0.193, respectively. This method is based on pixel-based classification, and hence, is slow and time-consuming.
 	
 	Regular sliding windows approach has been used to detect MAs by \citet{zhang2019detection}. First, contrast equalization preprocessing on the green channel is applied. Then, a deep neural network with a multilayer attention method is used for detection. They compared their work with faster RCNN that produces on average 300 proposals generated by region proposals network (RPN) and showed  that their method outperforms faster RCNN. They tested the method on IDRiD with average precision equal to 0.757 and sensitivity equal to 0.868, whereas with faster RCNN average precision equals 0.684. Also, \citet{chudzik2018microaneurysm2} proposed MAs detection method based on FCN and fine-tuned weights by freezing interleaved layers which restrict the weight changes and the trainable parameters. This method was tested on the e-ophtha with FROC equals 0.431 and the ROCh training dataset with FROC equals 0.298.
	
		The overview of the state-of-the-art red lesion detection methods given above shows that most of the methods deal with only one of the two red lesions, i.e., MAs or HMs. Only two methods focus on both MAs and HMs detection \cite{romero2019entropy,seoud2016red}, and most of the detection works focus on MAs or small red lesions due to candidates generation approaches used in the proposed methods that are unextended to HMs candidates generation \cite{orlando2018ensemble,long2020microaneurysms}. This is due to the features that are used to identify red lesions and focus on geometry metrics such as circularity and number of pixels \cite{orlando2018ensemble,wu2017automatic}. Moreover, a high number of MAs candidates used in detection \cite{eftekhari2019microaneurysm,chudzik2018microaneurysm1} leads to computation complexity problems. Moreover, though its MAs detection performance is good, the method by \citet{seoud2016red}, which detects both MAs and HMs,  suffers from the curse of dimensionality of shape and colors features and fails to detect all shape variations of HMs such as flame-shaped HM. Also, the method by \citet{romero2019entropy} reports overall red lesion detection performance without specifying the type of red lesion and has low computation efficiency due to a high number of candidates because they consider dark regions as candidates and the curse of dimensionality.		

 		\section{Proposed method} 
\begin{figure}
					\centering
					\includegraphics[width=13cm,height=7cm]{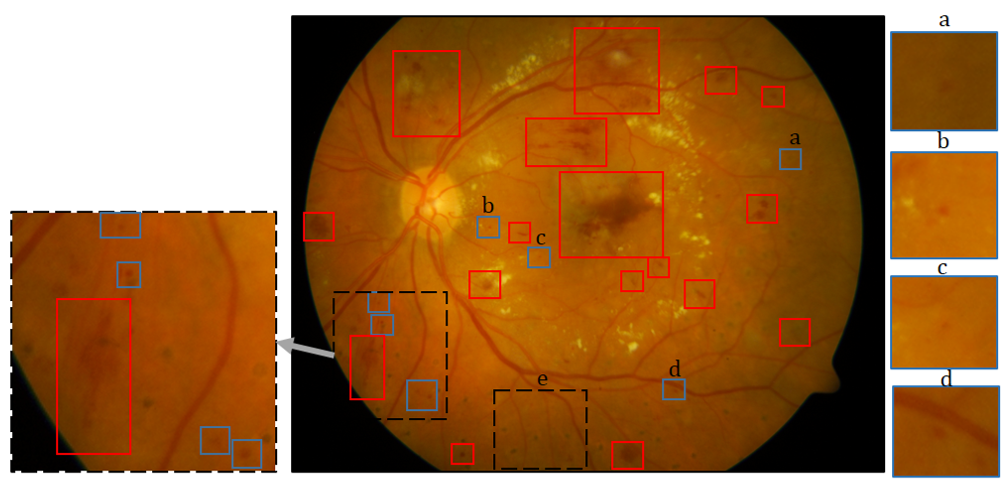}
					\caption{: A retinal fundus image (DiaretDB1: image005.png) showing red lesions: MAs (blue boxes, a-d), HMs (red boxes),  laser scars (e).}
					\label{fig:findings}
				\end{figure}

\begin{figure*}[]\centering 
		\includegraphics[width=20cm,height=10cm]{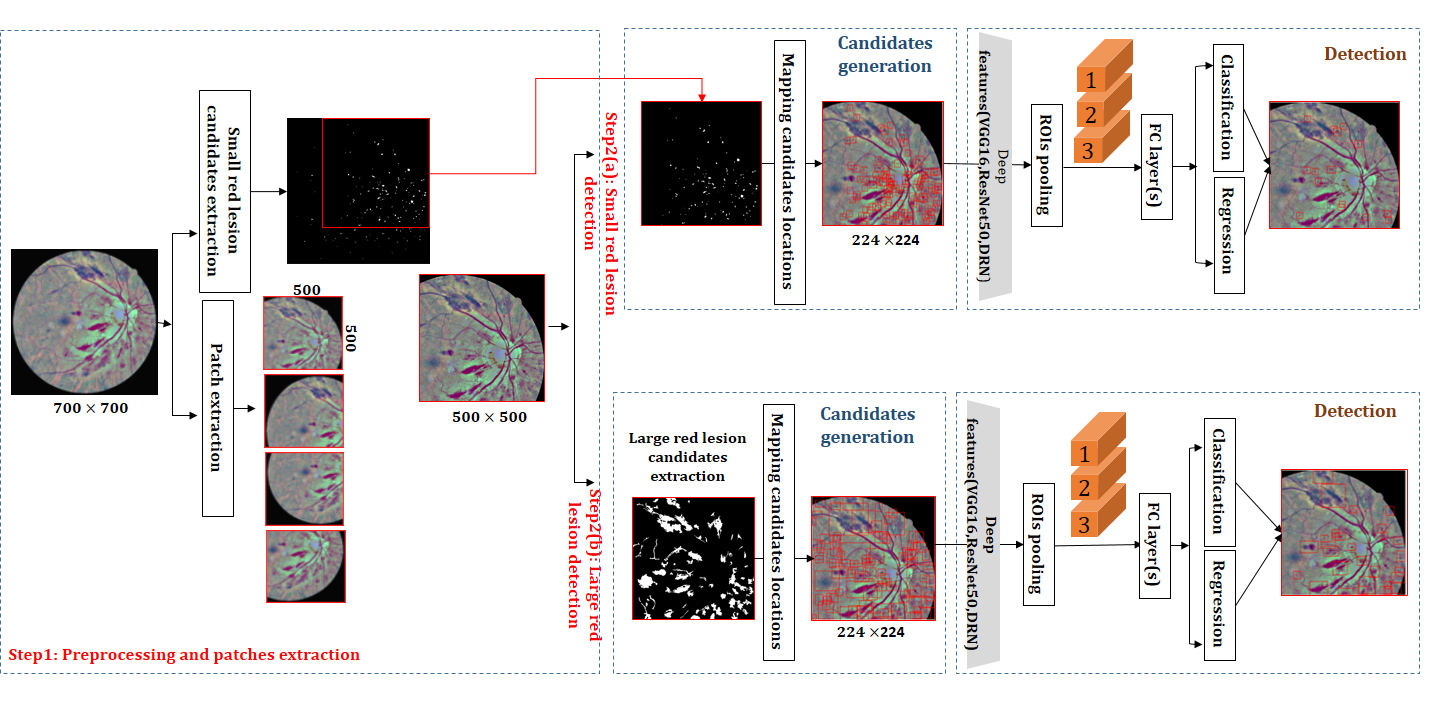}
		 \captionsetup[table]{font=small,skip=0pt}
		
		
		\caption{An overview of the proposed unified framework for red lesion detection on fundus images.}
			\label{fig:framework}
	\end{figure*}

 		The early clinical signs of DR are microaneurysms (MAs), which are small in size, i.e., less than 125 microns, as shown in Figure \ref{fig:findings}(blue boxes). As DR progresses, the number of MAs increases and the walls of some of them are broken, and bleeding causes hemorrhages (HMs) (see red boxes in Figure \ref{fig:findings}). Small HMs are similar to MAs but greater in size \cite{early1991grading}. Most of the published works do not detect small and large red lesions using a unified framework \cite{singh2018analysis}. Some focus on MAs detection only (\citet{wu2017automatic,eftekhari2019microaneurysm,chudzik2018microaneurysm1}). In contrast, others deal with the detection of MAs and small to medium red lesions which have circular shapes (\citet{orlando2018ensemble,adal2014automated}). Only a few works focus on both MAs and HMs (\citet{romero2019entropy,seoud2016red}). We propose a unified framework for the detection of small and large red lesions.
 		An overview of the proposed method is depicted in Figure \ref{fig:framework} . It consists of three main phases: preprocessing and patch extraction (Step1 in Figure \ref{fig:framework}), small red lesion detection (Step2 (a) in Figure \ref{fig:framework}), large red lesion detection (Step 2(b) in Figure \ref{fig:framework}), and postprocessing. The detail of preprocessing and patch extraction is presented in Section \ref{preprocessing}. Small and large red lesion detection involves two main operations, i.e., region candidates generation, and detection. Region candidates generation methods, which are different for small and large lesions, are presented in Section \ref{roi_gen}, but detection, which is similar for both, is presented in Section \ref{detection}. Finally, the postprocessing is described in Section \ref{postprocessing}.

\subsection{Preprocessing and Patches Extraction} \label{step1}	
	\subsubsection{Preprocessing}	\label{preprocessing}
\begin{figure}[H]
		\centering
	\includegraphics[width=.57\linewidth]{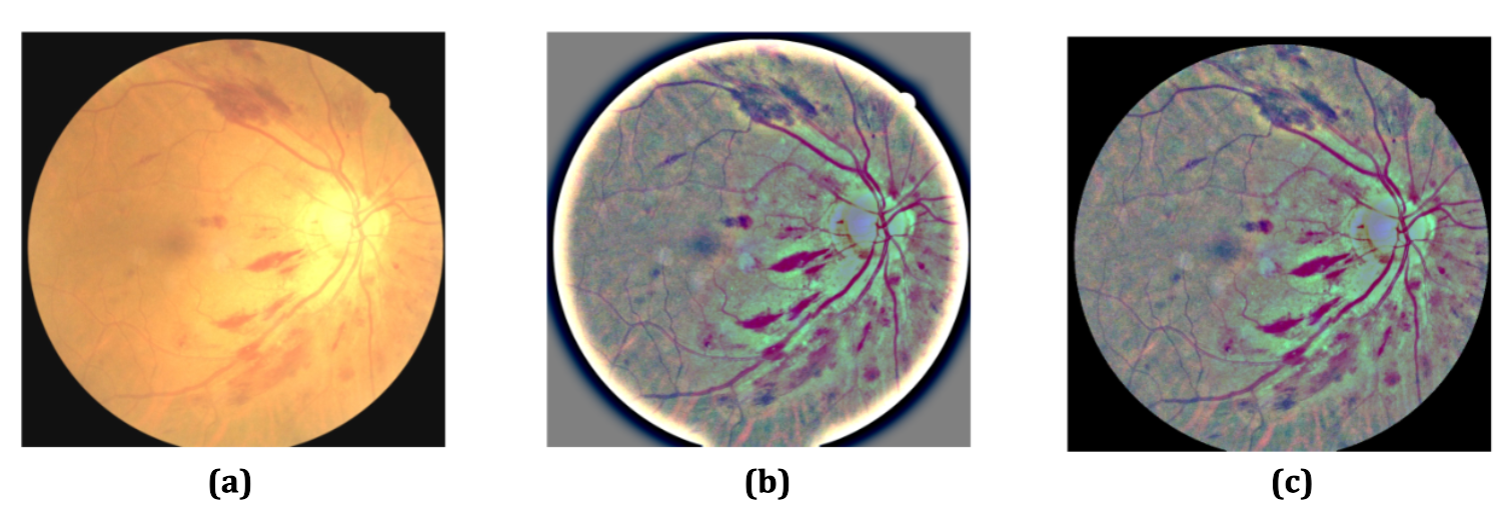}
	\caption{(a) Original RGB image, (b) CE preprocessed image without FOV mask, (c) CE preprocessed image $I_c$ with FOV mask.}
	\label{fig:screenshot004}
\end{figure}

	Fundus images usually suffer from the problem of illumination variation due to non-uniform diffusion of light in the retina. The curved surface of a retina is close to a sphere, which prevents uniform reflection of incident light and leads to hardly visible lesions \cite{asiri2019deep}. To overcome this problem, we use contrast equalization (CE). It is widely used as a preprocessing step in retinal fundus images to suppress noise, improve the contrast and tackle illumination variations \cite{walter2007automatic,orlando2018ensemble}. Moreover, we create FOV mask M to remove the black margin and then resize the image. We automatically generate the mask for each image by grouping the pixels into the mask and non-mask pixels using the red channel. In addition, this FOV mask is used to overcome CE’s undesired effects and bright artifacts that produce false detection on the FOV border and also hide potential lesions, as shown in Figure \ref{fig:screenshot004}, which happens because of the step edge along the border of FOV. This is solved by padding fundus image, which simulates a wider aperture and then considers only retina foreground using FOV mask \cite{soares2006retinal,orlando2018ensemble}.
		The mask is generated using the red channel. First, the contrast of the red channel is enhanced with power transform using power equal to 0.25. Then FOV mask is generated using fuzzy c-means (FCM) clustering algorithm \cite{bezdek1984fcm} with the number of clusters equal to 2.

In CE, after extracting mask $\mathbb{M}$, pixels in the green channel are padded to simulate a wider aperture around $\mathbb{M}$ \cite{soares2006retinal} by repeating $W = \frac{3}{30} \chi$ to ensure that pixels around the mask border are replaced by the mean of neighbors values which also include pixels inside the aperture. After that, each color band is equalized as follows:
  
 \begin{equation}
 I_{c}(i,j;\sigma) = (\alpha I(i,j)+ \tau Gaussian(i,j;\sigma)*I(i,j)+ \gamma)\times\mathbb{M}(i,j)
 \end{equation} 
 where $*$ is the convolution operation, the Gaussian filter has the standard deviation $\sigma =\chi/30$, $\alpha=4$, $\tau=-4$ and $\gamma=128$ \cite{orlando2018ensemble} and $\chi$ refers to the width in pixels of the field of view (FOV).

	\subsubsection{Patches extraction}
	Image size is a trade-off between speed and accuracy of detection\cite{huang2017speed}. Smaller images lead to faster detection. However, small ROIs might vanish through downsampling. On the other hand, large images lead to more accurate detection, but large input consumes more time; hence resizing the images is compulsory. To overcome these issues, instead of using the images actual resolution (e.g., $1,500\times1,152$ in DiaretDB1), first, we remove black margins based on the FOV mask and then resize the image into $700\times700$ since the minimum resolution of images is $650 \times 700$ in STARE dataset and also to accelerate processing for MAs candidates generation and for computational efficiency. Next, to overcome image downscaling issues and for more accurate detection, we divide the downscaled images into $2 \times 2$ overlapped patches $P_1, P_2, P_3$, and $P_4$, each of resolution $500 \times 500$, with total overlapped area 65\% from overall image to solve lesions cutting around macula and OD as depicted in Step1 of Figure \ref{fig:framework}.

\subsection{Region Candidates Generation} \label{roi_gen}

The main issue in candidates generation is the variation of properties of red lesions such as shape and size. MAs usually appear circular with a small number of pixels, and HMs have different red shades and irregular shapes such as circular, flame, and discrete lesions, as depicted in Figure \ref{fig:findings}. Usually, extracting small red lesion candidates is easier than large candidates \cite{orlando2018ensemble,wu2017automatic,long2020microaneurysms}. Using the same method, such as morphological operations, to generate small and large red lesions candidates does not work because blood vessels are retrieved as candidates as well \cite{orlando2018ensemble}. A brute force method to generate small and large red lesions candidates together is selective search \cite{tang2017selective}. However, such methods generate $\sim$2,000 candidates per image. In the proposed method, we extract a small number of candidates in two different ways with an average of 28 candidates for MAs and 75 candidates for HMs as shown in Table \ref{train}, and then feed them into two different streams based on candidates type to make the detection accurate and fast.

\subsubsection{Small red lesion candidates}

\begin{figure}[H]
	\centering
	\includegraphics[width=1\linewidth]{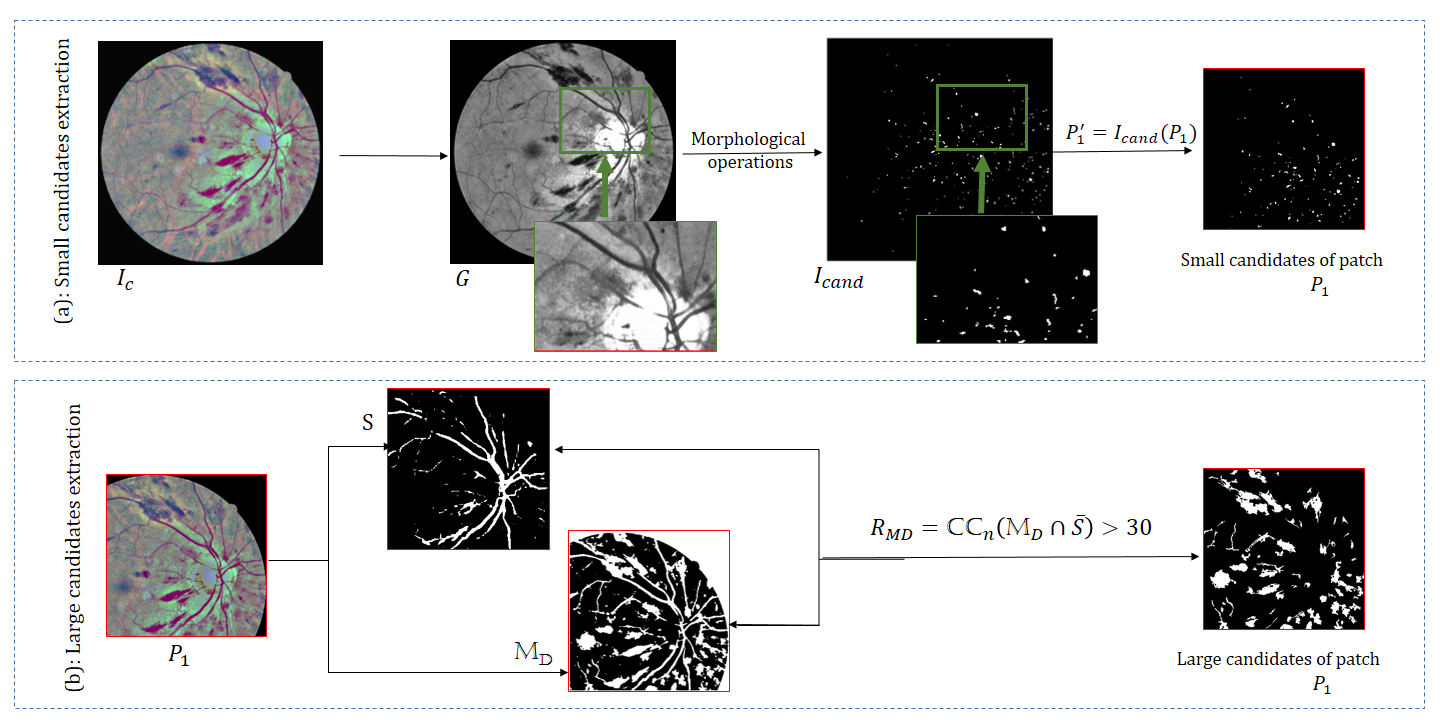}
	\caption{An overview of red lesions candidates extraction. (a): small red lesion candidates, (b): large red lesion candidates.}
	\label{fig:red_lesions_cands}
\end{figure}

For small red lesion candidates extraction, we used the method proposed by \citet{orlando2018ensemble}, which is an unsupervised method based on a set of simple morphological operations. First, the green channel  $G$ of the enhanced image $I_c$ is extracted. After that, an $r$-polynomial transformation \cite{walter2007automatic} is applied on $G$ to overcome the issue of uneven background illumination, which hides lesions in dark areas. Then, $G$ is filtered with a Gaussian filter to reduce noise. Next,  images are obtained from $G$ by using morphological closing operation with line structuring elements of lengths $l\in \{3,6,9,...,60\}$ and angles spanning from $0^{\circ} to 180^{\circ}$, and $I_{cand}^l = I_{closed}^l - G$  is computed for each $l$. Then each  $I_{closed}^l$ is thresholded in such a way that the number of lesion candidates left in $I_{closed}^l$ is less or equal to $K=120$ and $I_{cand} = \bigcup\limits_{l\in \{3,6,9,...,60\}} I_{cand}^l$ is obtained as shown in Figure \ref{fig:red_lesions_cands}(a). Finally, the very small connected components with less than 5 pixels are ignored since they are related to noise or other background particles. The detail can be found in \cite{orlando2018ensemble}. 

\subsubsection{Large red lesions candidates}
Unlike MAs and small red lesions, HMs and large red lesions have different properties such as shapes and sizes,  as shown in Figure \ref{fig:findings}(red boxes). If the same method, which is used to create candidates of MAs, is employed to extract candidates of large red lesions, then blood vessels are extracted as candidates. Also, the method becomes very slow because of the morphological closing operation, which involves line structuring elements of very large length. Large red lesions appear as dark regions in the green component of a patch, and they can be extracted by segmentation using thresholding. However, this approach extracts large red lesions and dark regions such as blood vessels, which cause a large number of false positives. The solution to this issue is to remove blood vessels. After removing blood vessels, the remaining dark regions correspond to either red lesions or disentangled vessel segments or fovea. To remove blood vessels, first, we segment blood vessels then remove them from the patch.

In general, removing retinal blood vessels is a frequently applied step in detecting pathologies on fundus images. However, it is not a straightforward operation due to their low contrast, variations in their morphology against the noisy background, and the presence of pathologies like MAs and HMs \cite{lam2008novel}. Several vessels segmentation techniques have been proposed \cite{maji2015deep,liskowski2016segmenting,maninis2016deep}. \citet{jiang2019automatic} segmented the blood vessels using a fully convolutional network (FCN) \cite{long2015fully}. We adopt this method employed with FCN-8. However, unlike \citet{jiang2019automatic}; we use the pre-trained VGG-16  model with three channels instead of a single channel to add more contextual information. Before the segmentation of vessels using FCN-8, an image is preprocessed using CE and divided into overlapping patches of size $500 \times 500$ pixels each. For FCN-8, we used the pre-trained VGG-16 model as the backbone and fine-tuned it using two benchmark public datasets with blood vessels annotations, i.e., DRIVE \cite{drive} and STARE \cite{hoover2000locating}. For fine-tuning, we extracted 300,125 patches of size $500\times500$ pixels with their ground truth using the annotations of the databases. Since the number of annotated images is limited (see Table \ref{train}), to enhance the number of patches for training FCN-8, we extracted patches such that each patch was centered on a random vessel pixel. We used stochastic gradient descent (SGD) with a momentum of 0.9 and a small learning rate (i.e., $\eta = 0.0001$) and a batch size of 20 patches.

\begin{figure}
\centering
\includegraphics[width=.5\linewidth]{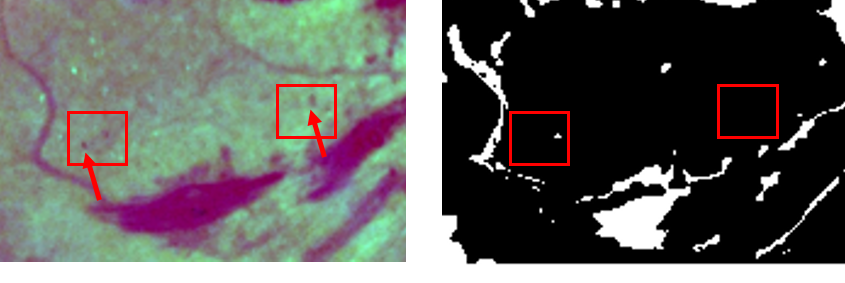}
\caption{ Small MAs candidates are eliminated after vessels segmentation.}
\label{fig:image}
\end{figure}

After training FCN-8, vessels pixels are segmented from patch $P_i$, and binary mask S consisting of segmented vessels is obtained. After that, the green component of patch $P_i$, namely $G_i$, is segmented using thresholding with threshold $D\leq0.45$, and another binary mask $\mathbb{M}_D$ is obtained, which contains all dark regions, including blood vessels. For removing blood vessels from $\mathbb{M}_D$ , its intersection with the complement of $S$ is computed, i.e., $R_{\mathbb{M}_D} = \mathbb{M}_D \cap \overline{S}$, where $\overline{S}$ is the complement of $S$ and $\cap$ is an intersection operation. The $R_{\mathbb{M}_D}$ contains only large red lesions candidates and noises. The connected components with less than 30 pixels are discarded for removing noises because connected components with less than 30 pixels are either MAs or other noises. For this purpose, we apply the operation $\mathbb{CC}_n \cap \overline{S} >30$ where $\mathbb{CC}$ stands for connected components algorithm; this operation keeps the connected components with pixels greater than 30 pixels, which are large red lesion candidates. This method is not suitable for MA candidates extraction. The blood vessels segmentation process using VGG-16 performs downsampling by the rate of 32, and any region less than $32\times32$ region vanishes. So MAs are eliminated because the average size of MAs is $21\times21$ as shown in Figure \ref{fig:image}.


\subsection{Detection} \label{detection}
The extracted small and large red lesion candidates are potential reigns of interests (ROIs), and each ROI is specified by four values $(r,c,h,w)$, where $(r,c)$ are the coordinates of the center, and the height and width of its bounding box, respectively. After the extraction of ROIs, the next step is to reduce the false positives and fine-tune the bounding boxes of the true ROIs. The false-positive reduction is a binary classification problem (red lesion, no red lesion), and fine-tuning the bounding boxes is a regression problem. Following the strategy of Fast-RCNN \cite{girshick2015fast}, we solve these two problems using VGG-16 as the backbone model. In this approach, each patch $P_i$ and red lesion candidates are passed to VGG-16 to extract features of each ROI, which are further passed to two subnets for classification and regression. There are two issues: (i) the dimensions of the features of different ROIs are different, but each subnet takes the input of fixed dimension, (ii) the texture patterns of small and large red lesions are different, and the same backbone CNN for feature Extraction does not accurately encode the characteristics of each type of lesion. The first problem is solved by using ROI pooling \cite{girshick2015fast}, which maps features corresponding to ROIs of different sizes to a fixed size. We use two streams to tackle the second problem, one for small red lesions and the other for large red lesions, as shown in Figure \ref{fig:framework}. We used pre-trained VGG-16 because it has been widely used in many applications and is suitable for texture representation \cite{li2017scale}; it is a reasonable choice for representing small and large red lesions. The ROI pooling layer replaces the last max-pooling layer in VGG-16 to pool the feature maps of each candidate into fixed resolution and retrieve features of all ROIs at once. Also, the final fully connected layer and softmax layer in VGG-16 are replaced with two fully connected layers: classification and regression layers. Given an ROI feature vector $r$ with ground truth offset $v$, the classification layer processes $r$ and yields the probability vector $p_r = [p_{r_{dr}},1-p_{r_{dr}}]$ where $p_{r_{dr}}$ is the probability of $r$ a being red lesion (i.e., MA or HM based on stream). A box regression layer provides a finer bounding box location.

\begin{table*}[ht]
	\centering
	\caption{Datasets used in training}
 
	\begin{tabular}{|l|l|l|l|l|l|l|l|}
		\hline
		\multirow{2}{*}{Dataset} & \multirow{2}{*}{Lesion types} & \multirow{2}{*}{\begin{tabular}[c]{@{}l@{}}No. images-\\ patches\end{tabular}} & \multirow{2}{*}{\begin{tabular}[c]{@{}l@{}}No. ground\\ truth boxes\end{tabular}} & \multirow{2}{*}{No. candidates} & \multicolumn{3}{l|}{\begin{tabular}[c]{@{}l@{}}Avg. No. \\ candidates\end{tabular}} \\ \cline{6-8} 
		&                               &                                                                                &                                                                                   &                                 & Proposed                      & SS                        & RPN                     \\ \hline
		DiaretDB1 / DDR/ IDRiD   & MA                            & 651- 13000                                                                     & 49,396                                                                            & 358,272                         & 28                            & 2000                      & 300                     \\ \hline
		DiaretDB1 / DDR/ IDRiD   & HM                            & 707- 29,732                                                                    & 416,828                                                                           & 2,218,167                       & 75                            & 2000                      & 300                     \\ \hline
		DRIVE/ STARE             & Blood vessels                 & 80- 300,125                                                                    & -                                                                                 & -                               & -                             & -                         & -                       \\ \hline
	\end{tabular}
 
\label{train}
\end{table*}

\subsection{Postprocessing} \label{postprocessing}

In this step, for evaluation and representation purpose, we merge all the patches retrieved from MA and HM branches into one single image. In total, eight patches are obtained as follows: $[P^{ma}_1, P^{ma}_2, P^{ma}_3, P^{ma}_4]$ received from MAs detection branch and $[P^{hm}_1, P^{hm}_2, P^{hm}_3, P^{hm}_4]$ received from HMs detection branch  with a size $500 \times 500$ of each patch. Every two patches of same number are merged into one patch using $P_j  (x^i,y^i )=max(⁡P^{ma}_j(x^i,y^i),P^{hm}_j(x^i,y^i))$. After that, we localize these merged patches into $700 \times 700$ empty image $I_{output}$. For the overlapped location where we have union of the 2 patches the output is obtained using: $I_{output}(x^i,y^i) = max(P_1(x^i,y^i),P_2(x^i,y^i))$ and then for 4 overlapped patches $I_{output}(x^i,y^i) = max(P_3(x^i,y^i),P_4(x^i,y^i),I_{output}(x^i,y^i))$.

\section{Training the System}
Due to the unavailability of huge annotated fundus images, we extracted patches from different datasets, as shown in Table \ref{train}. Patches based augmentation approach was employed to increase the number of training patches. We used rotation with angles in [$-45^{\circ},79^{\circ}, 90^{\circ}$] and nearest-neighbor interpolation.
  
For small red lesion (MAs) detection, we extracted 13,000 patches from 651 images collected from DiaretDB1 \cite{dia}, IDRiD\cite{idrid} and DDR \cite{li2019diagnostic} databases. There are 49,396 MAs (ground truth), but the small red lesion candidates extraction method found 358,272 MAs candidates in these patches, with 28 candidates for each patch on average.
For large red lesion (HM) detection, we extracted 29,732 patches from 707 images collected from DiaretDB1 \cite{dia}, IDRiD\cite{idrid} and DDR \cite{li2019diagnostic} databases. These patches contain 416,828 HMs (ground truth), but the large red lesion extraction algorithm found  2,218,167 HMs candidates with 75 candidates for each patch on average. 
A candidate is considered a positive example if it has an IOU ratio greater than 0.5 with one ground-truth box, and otherwise, it is a negative sample. The positive and negative candidates are sampled into mini-batches to speed up training and overcome memory limitations. Non-maximum suppression (NMS) is used to manage duplicated and overlapped boxes using their IOU, which controls the increase of false positives.

For training, we used stochastic gradient descent (SGD). To take advantage of feature sharing during training, we sample $N$ images and then $R$ ROIs from $N$ images; in our experiments, we used $N = 2$ and $R = 64$ in each mini-batch. To overcome the overfitting, we added two dropout layers after $FC6$ and $FC7$ layers of VGG-16. We empirically set the dropout rate $drop_{MA}$ = 0.8 for dropout layers in MA stream and MA stream and $drop_{HM}$ = 0.7 for the HM stream.

We used multi-task loss $L$ for each ROI labeled as a red lesion (i.e., MA or HM) or background jointly train the classification and bounding box regression nets. The label of a true red lesion is set $u = 1$, whereas that of a non-lesion $u = 0$ and predicted offset $t^u=[t_r^u,t_c^u,t_h^u,t_w^u]$ and ground truth offsets $v = [v_r,v_c,v_h,v_w]$ \cite{zhao2019object}. The joint loss $L$ of classification and regression is calculated as follows:

\begin{equation}
L=L_{cls}(p_r,u)+ 1[u\ge 1]L_{loc}(t^{u},v)
\end{equation}
where $p_r$ is the predicted confidence score, $L_{cls}$ and $L_{loc}$are cross-entropy losses \cite{hwang2016self,zhao2019object}, which are defined as follows: 
\begin{equation}
L_{cls}(p_r,u) = -log p_{r_u} ; 
\end{equation}
where $p_{r_u}$ is the probability of true class $u$.
\begin{equation}
L_{loc}(t^{u}, v) = \sum_{i\in r,c,h,w} smooth_{L_1}(t^{u}_i - {v_i})
\end{equation}

\begin{equation}
smooth_{L_1}(x) = 
\begin{cases} 0.5x^2 \ if \ |x| \ < \ 1 \\ 
|x| \  - \  0.5 \ otherwise 
\end{cases}
\end{equation}

The  $1[u\ge 1]$ equals 1 when $u\ge 1$ for red lesion, and 0 otherwise for background proposal \cite{zhao2019object}.

\section{Evaluation Protocol}
In the proposed method, we use datasets with annotations for blood vessels \cite{drive,hoover2000locating} commonly used for vessel segmentation tasks and red lesions annotation \cite{kalviainen2007diaretdb1,optha,porwal2020idrid,niemeijer2010retinopathy} used for red lesions detection tasks. Fundus images in these datasets were gathered by different fundus cameras with different degrees, quality and conditions, and resolution, as shown in Table \ref{train_test}.
Some of the red lesions datasets include both MAs and HMs annotations such as DiaretDB1\cite{kalviainen2007diaretdb1}, IDRiD\cite{idrid}, and DDR \cite{li2019diagnostic} and some have only MAs annotation such as in e-ophtha\cite{optha} and ROCh \cite{niemeijer2010retinopathy}. This leads to distribution imbalance of red lesions among these datasets, as shown in Table \ref{train_test}; for example, in DiaretDB1\cite{dia}, not all pathological images have MAs and HMs (i.e., out of 89 images, 74 have MAs annotations, and 53 images are HMs annotations). Also, for the e-ophtha dataset, only MAs are highlighted in the ground truth, and HMs are ignored.  
We selected the state-of-the-art red-lesion detection algorithms for comparison. Deep learning techniques were implemented in MATLAB R2017a, using Matconvnet \cite{vedaldi2015matconvnet}.The hardware specifications of our laptop includes NVIDIA GeForce GTX 1070 GPU, Intel Core i7-7700HQ CPU@ 2.80 GHz processor and 32.0 GB of RAM. All training and testing were performed in the same hardware environment. 

\subsection{Datasets}

\subsubsection{Blood vessels segmentation datasets}
Digital retinal images for vessel extraction (DRIVE) \cite{drive} was obtained from the Netherlands diabetes retinopathy screening program using CR5 non-mydriatic 3CCD camera. It is focused on vascular segmentation in fundus images and provides pixel-level annotation. The DR screening of 400 diabetic patients between 25-90 years of age was done; 40 random fundus images with the resolution of $584\times565$ were selected; 33 showed no signs of diabetic retinopathy, and 7 showed signs of moderate diabetic retinopathy.

Structured analysis of the retina (STARE) \cite{hoover2000locating} contains 40 retinal fundus images with ground truth of blood vessels. The images have image-level annotations of 13 eye diseases and pixel-level annotations of blood vessels and optic nerve. The resolution of each image is $605\times700$, with 24 bits per pixel.

\subsubsection{Red lesions detection datasets}

DiaretDB1 \cite{kalviainen2007diaretdb1,dia} consists of 89 color fundus images, 84 of which contain at least one non-proliferative sign of diabetic retinopathy such as MAs, HMs, and EX, and 5 are normal and have no signs of DR according to all four experts involved in the assessment. The images were taken using the 50$^{\circ}$ digital field vision fundus camera with a resolution of $1,500\times 1,152$.

\begin{figure}
\centering
\includegraphics[width=0.7\linewidth]{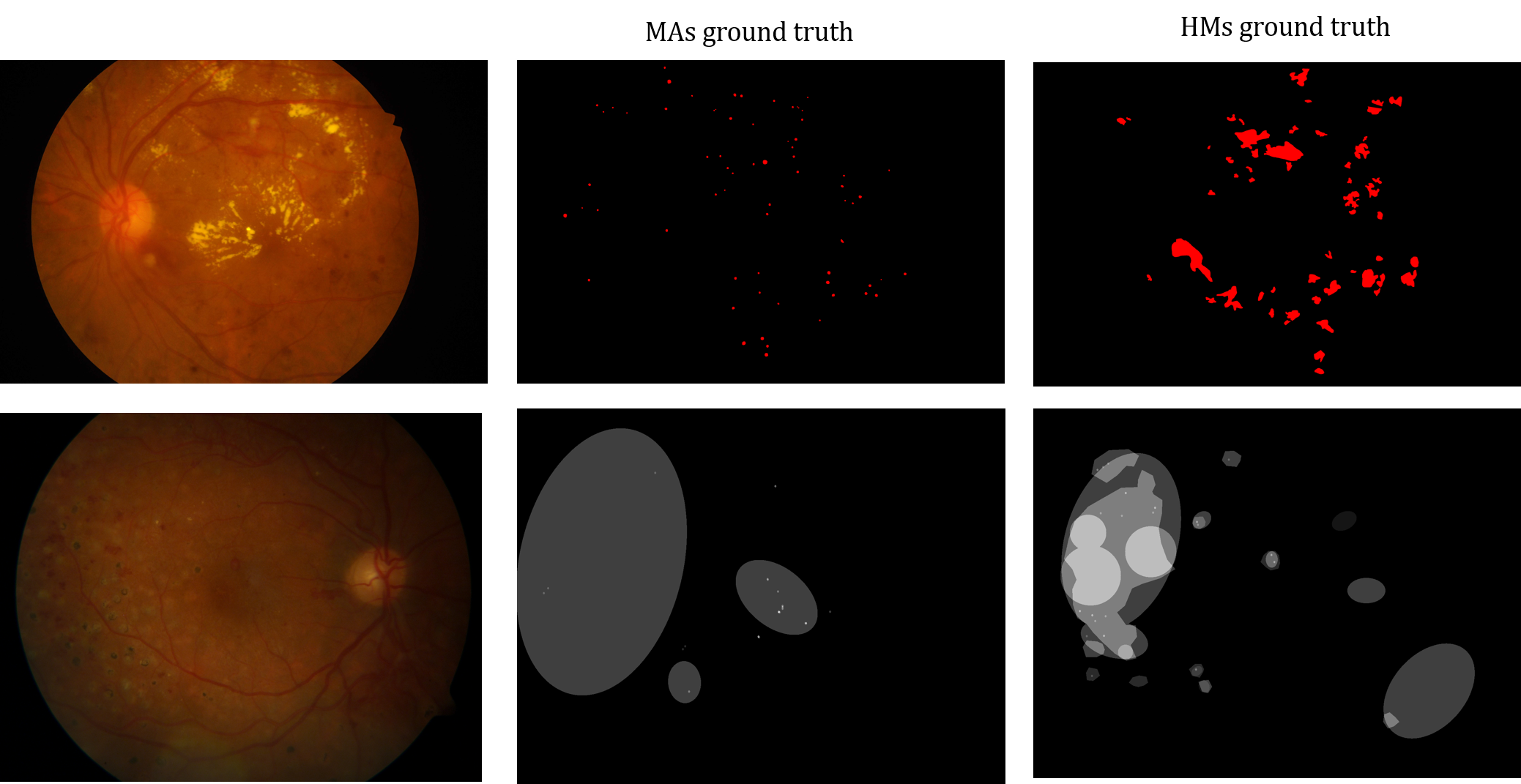}
\caption{An example of image annotations: first row: IDRiD\_25 from IDRiD dataset and related HMs and MAs annotations showing exact location of the lesions. Second row: image026 from DiaretDB1 and related MAs and HMs annotations showing confidence degree of red lesions.}
\label{fig:screenshot012_hm}
\end{figure}

e-ophtha \cite{optha} is a dataset of color fundus images dedicated to DR research funded by the French research agency. The images contain the ground truth of MAs and EXs provided by two ophthalmologists. e-ophtha consists of two databases called e-ophtha-MA and e-ophtha-EX (EXudates). e-ophtha-EX includes 47 images with exudates and 35 normal images with no lesion. On the other hand, e-ophtha-MA contains 148 images with 1306 MAs regions and 233 healthy images. In this research we use, e-ophtha-MA and refer to it e-ophtha in the rest of this paper.

Indian Diabetic Retinopathy Image Dataset (IDRiD) dataset \cite{porwal2020idrid} consists of 516 color fundus images with signs of DR and DME taken from an Indian population. The images were captured and verified from an eye clinic located in Nanded in India using a Kowa VX-10 alpha digital fundus camera with 50$^{\circ}$ FOV, and all are centered near to macula with the resolution of $4288\times2848$. For pixel-level annotation, binary masks in tif format of 54 images with MAs and 53 images with HMs are provided.

Dataset of Diabetic Retinopathy (DDR) \cite{li2019diagnostic} is a general purposes dataset containing poor quality images to reduce overfitting.  It was collected from different imaging devices and hospitals across China. It includes 13,673 fundus images with image-level annotations where 6266 are healthy, 6256 show DR signs, and 1151 are ungradable. Among them, 757 images with pixel-level and bounding box level annotations are provided. It is known to be the largest pixel-level annotation dataset. The image-level annotations are as follows: no DR: 6266, mild: 630, severe: 236, proliferative: 913, ungradable: 1151. Pixel-level annotations include 570 images for MA,  601 for HM, 239 for soft EX, and 486 for hard EX.

Retinopathy Online Challenge (ROCh) is dedicated to MA detection in fundus image with $45^{\circ}$. It includes publicly available 50 images for training and 50 images for testing, and only ground truth annotation is available for the training dataset in an XML file format \cite{niemeijer2010retinopathy}. It consists of images with three resolutions: $768 \times 576, 1058 \times 1061$, and $1389 \times 1383$. The images were captured using either a Topcon NW 100, a Topcon NW 200, or a Canon CR5-45NM.

\paragraph{Private Dataset}
Private dataset The fundus images were collected from the Diabetes Center of King Abdulaziz University Hospital in Saudi Arabia (DCKAU).  The 3D OCT MAESTRO with color non-mydriasis fundus photography with $45^{\circ}$ was used to capture the fundus image of each eye, the one centered on the optic disc and the other centered on the macula. One expert annotated the images and contained 37 images with moderate DR and 15 images with severe DR. The camera produces two adjacent photos, one colored and the second is the green channel,  which causes poor quality images.

\subsection{Evaluation Procedures}
We extend the candidate bounding boxes by adding 10 pixels in all directions to include more feature information of the lesion's shape, texture and context. Also, we normalize all patches produced from e-ophtha and DiaretDB1 datasets using their overall mean and the current patch mean for ROCh and private datasets. We use similar training settings for all datasets; for MA, we choose $\theta_{MA} \geq 0.1$ and for HM we choose $\theta_{HM} \geq 0.1$. For NMS, for MA we set $NMS_{MA} = 0.3$ and for HM $NMS_{HM} = 0.5$.  

We test the proposed system on 543 images taken from 4 datasets: e-ophtha (380 images), DiaretDB1 (61 images), ROCh (50 images), and private dataset (37 moderate DR and 15 severe DR images), as shown in Table \ref{train_test}. We use two different evaluation metrics one for per-lesions and one for image screening. For per-lesion detection, we use the performance metrics that are commonly used to assess and evaluate the overall red lesions detection. A standard metric for measuring the performance of algorithms is free-response ROC (FROC) \cite{orlando2018ensemble,seoud2016red,niemeijer2010retinopathy}, which plots per lesion sensitivity under specific number of FPIs against the average number of false-positive detection per image (FPI). It shows the model’s ability to detect true red lesions in all test images \cite{orlando2018ensemble,seoud2016red}. To obtain the final sensitivity score and compare it with other methods, we calculate the competition metric (CPM) or FROC score proposed in the Retinopathy Online Challenge \cite{niemeijer2010retinopathy,orlando2018ensemble}. This metric averages the sensitivities at specific reference FPI points $\in \{1/8, 1/4, 1/2, 1, 2, 4, 8\}$. For image-level detection, to determine an image $I$ as DR/ no DR based on red lesions existence, we followed the same procedure used by \cite{seoud2016red,orlando2018ensemble}. Given $r_j$ the feature vector of ROI and the output probability $p_{r_j}$, then the $p(I)$ of the image $I$ is obtained as follows: $p(I) =  max_i(p_{r_j}|u = 1)$. Also we report sensitivity and specificity defined as:

\begin{equation} \label{eq:sens}
Sensitivity=\frac{TP}{TP+FN}
\end{equation} \label{eq:spec}
\begin{equation}
Specificity=\frac{TN}{TN+FP}
\end{equation}

\section{Results}

We quantitatively assess the model’s ability to detect both MAs and HMs simultaneously at multiple scales. The method is evaluated for lesion-level detection when delineation and annotation of red lesions are provided with the dataset. We carried out several experiments to evaluate the effectiveness of the proposed approach using three public datasets (testing images of DiaretDB1 and e-ophtha), the training set of ROCh, and one private dataset, all having pixel-level annotations. the proposed method versus other methods on e-ophtha, ROCh, DiaretDB1-MA, and DiaretDB1-HM datasets.
Tables \ref{cpmema} ,\ref{cpmdma},\ref{cpmdmaroch}, and \ref{cpmdhm} show a comparison of sensitivity between the proposed method and other algorithms \cite{wu2017automatic,long2020microaneurysms,orlando2018ensemble,seoud2016red,eftekhari2019microaneurysm,adal2014automated,chudzik2018microaneurysm1,chudzik2018microaneurysm2,zhang2010detection,javidi2017vessel} at different FPI and FROC score on e-ophtha, ROCh, DiaretDB1-MA, and DiaretDB1-HM datasets respectively.

\begin{table}[]
	\caption{Datasets used in training and testing red lesion}
	\begin{tabular}{|l|l|l|l|l|}
		\hline
		Dataset   & \#images                                                             & Resolution                                                              & Format                                                         & Task                                                                                                                            \\ \hline
		DiaretDB1 & \begin{tabular}[c]{@{}l@{}}26 (training)\\ 61 (testing)\end{tabular} & 1500$\times$1152                                                               & PNG                                                            & \begin{tabular}[c]{@{}l@{}}MA: 73 (26 training/47 testing)\\ HM: 53 (23 training/30 testing)\\ normal: (3 testing)\end{tabular} \\ \hline
		e-ophtha  & 380 (testing)                                                        & \begin{tabular}[c]{@{}l@{}}2544$\times$1696\\ 1440$\times$960\end{tabular}            & \begin{tabular}[c]{@{}l@{}}images: jpeg\\ GT: png\end{tabular} & \begin{tabular}[c]{@{}l@{}}MA: 148\\ normal: 233\end{tabular}                                                                   \\ \hline
		IDRiD     & 54 (training)                                                        & 4288$\times$2848                                                               & \begin{tabular}[c]{@{}l@{}}images: jpeg\\ GT: tif\end{tabular} & \begin{tabular}[c]{@{}l@{}}MA: 54\\ HM: 53\end{tabular}                                                                         \\ \hline
		DDR       & 757 (training)                                                       & varies                                                                  & \begin{tabular}[c]{@{}l@{}}images: jpeg\\ GT: tif\end{tabular} & \begin{tabular}[c]{@{}l@{}}MA: 570\\ HM: 601\end{tabular}                                                                       \\ \hline
		ROCh      & 50 (testing)                                                         & \begin{tabular}[c]{@{}l@{}}768$\times$576\\ 1058$\times$1061\\ 1389$\times$1383\end{tabular} & \begin{tabular}[c]{@{}l@{}}images: jpeg\\ GT:XML\end{tabular}  & MA: 50                                                                                                                          \\ \hline
		Private   & 52 (testing)                                                         & varies                                                                  & \begin{tabular}[c]{@{}l@{}}images: PDF\\ GT: PDF\end{tabular}  & \begin{tabular}[c]{@{}l@{}}37 moderate DR (MA+HM)\\ 15 severe DR images(MA+HM)\end{tabular}                                     \\ \hline
	\end{tabular}
\label{train_test}
\end{table}

\subsection{Lesion Level Detection Results}

For each lesion type, MAs and HMs, different experiments were performed to evaluate per lesion detection on different datasets. We used FROC scores, sensitivity and specificity to evaluate the performance at $\theta = 0.5, 0.3,0.1$. For FROC metric, we used 100 per-lesion sensitivity values (from 0 to 1) and the average number of false positives per image (FPI) retrieved for logarithmically spaced FPI thresholded based on per-lesion output probabilities. 
 
 For e-ophtha, we used only MA detection branch; the method reports the best $FROC_{MA} = 0.4537$ at $\theta=.1$ with overall sensitivity of 0.8157. 
 For DiaretDB1, two experiments were conducted, one for MA and the other for HM detection. For DiaretDB1-MA, the method reports $FROC_{MA_\theta=.1} = 0.2758$ with good sensitivity which reaches 0.8589. However, FROC score is minimum among other datasets due to dataset annotation issues discussed in \cite{asiri2019deep} such as geometrical annotation instead of pixel wise annotation as shown in Figure \ref{fig:screenshot012_hm}. Also some small red lesions were annotated as hemorrhages not MAs. 

On DiaretDB1-HM, the method gave  $FROC score_{HM_{\theta=.1}} = 0.4353, SN_{\theta=.1}=0.7552$, $SP_{\theta=.5} = 0.9633$, which is less than $FROC score_{HM}  = 5044$ achieved by \cite{orlando2018ensemble}.

On ROCh dataset, it achieved $FROC_{MA_\theta=.1} = 0.3461$ with SN = 0.6595, the method by \citet{chudzik2018microaneurysm2} got $FROC score = 0.298$. The low sensitivity is due noise, compression artifacts, and low resolution. We used the same evaluation protocol as was used in \citet{chudzik2018microaneurysm2}.  

Tables \ref{cpmema} and \ref{cpmdma} show that MAs detection on the e-ophtha and ROCh datasets is better than on the DiartDB1 dataset. It is because in DIaretDB1 MAs ground truth annotations are not always highlighted by their size and shape appearance as shown in Figure \ref{fig:screenshot012_hm}. In Figure \ref{fig:screenshot012_hm}, for example, the first row shows IDRiD ground truth annotation, any red lesion is highlighted as exact MAs or HMs; in contrast, the second row taken from DiaretDB1, the confidence degree is used for lesions annotation which leads to wrong locations. In addition, in e-ophtha, the number of normal images is greater than that of the images with MAs compared to DiaretDB1-MA, as shown in Table \ref{train_test}.

As discussed above, we notice that FROC and sensitivity increase with low $\theta$ values in contrast to specificity as presented in Tables \ref{cpmdma}, \ref{cpmdmaroch}, and\ref{cpmema}. This is due to increasing the number of true positives and false positives.

\begin{figure}.

	\centering
		\includegraphics[width=1\linewidth]{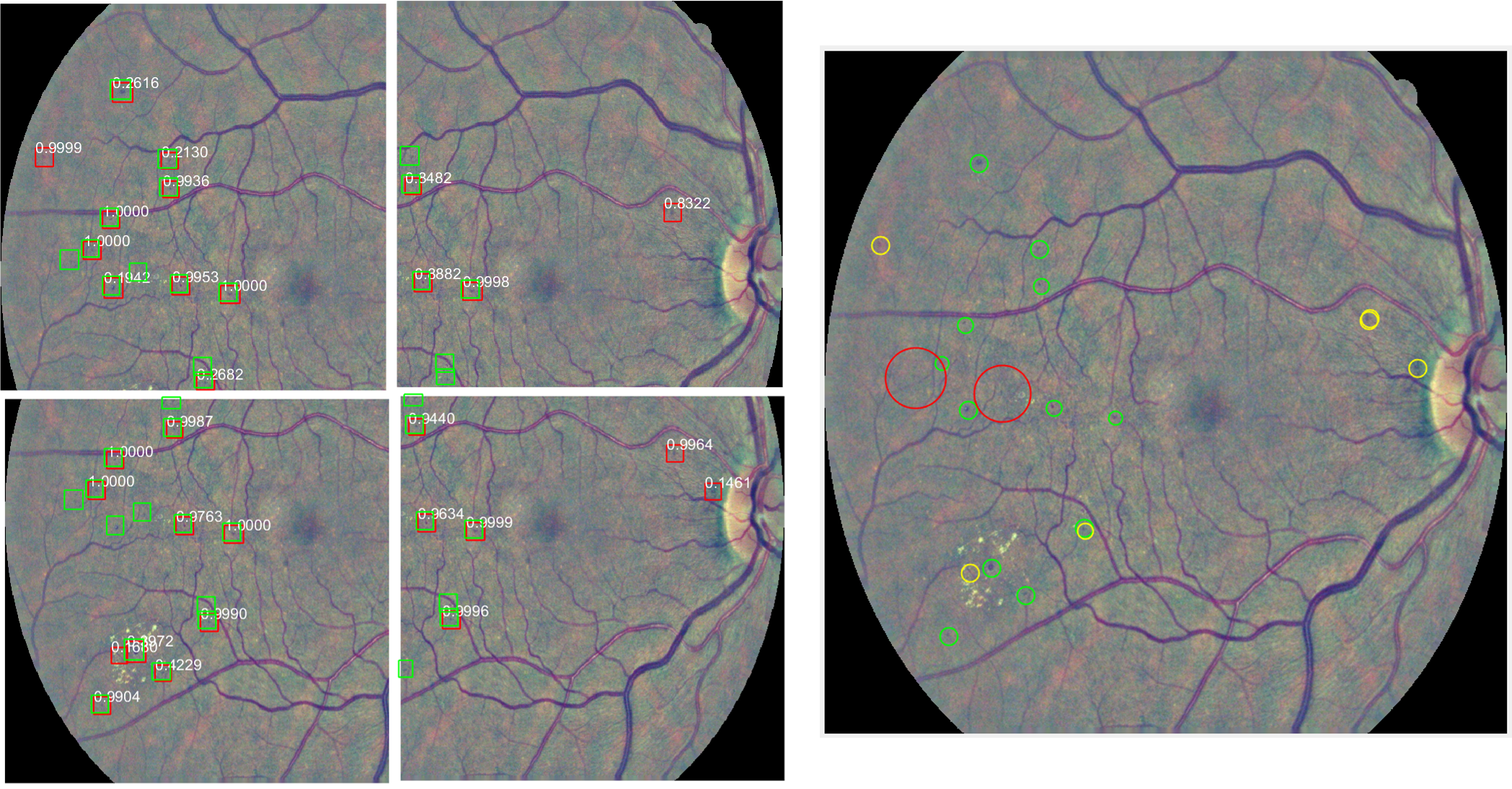}
		\caption{MAs detection results on the overlapped patches of C0003164 from the e-ophtha test set. At $\theta_{MA} \geq 0.1$ and $NMS_{MA} \geq 0.3$. Left: green rectangles show the detected TP MAs. Right: green circles show TP, yellow circles show FP and red ones show FN.} 
		\label{fig:results_2_a}
	\end{figure}

\begin{figure}	
	\centering
	\includegraphics[width=1\linewidth]{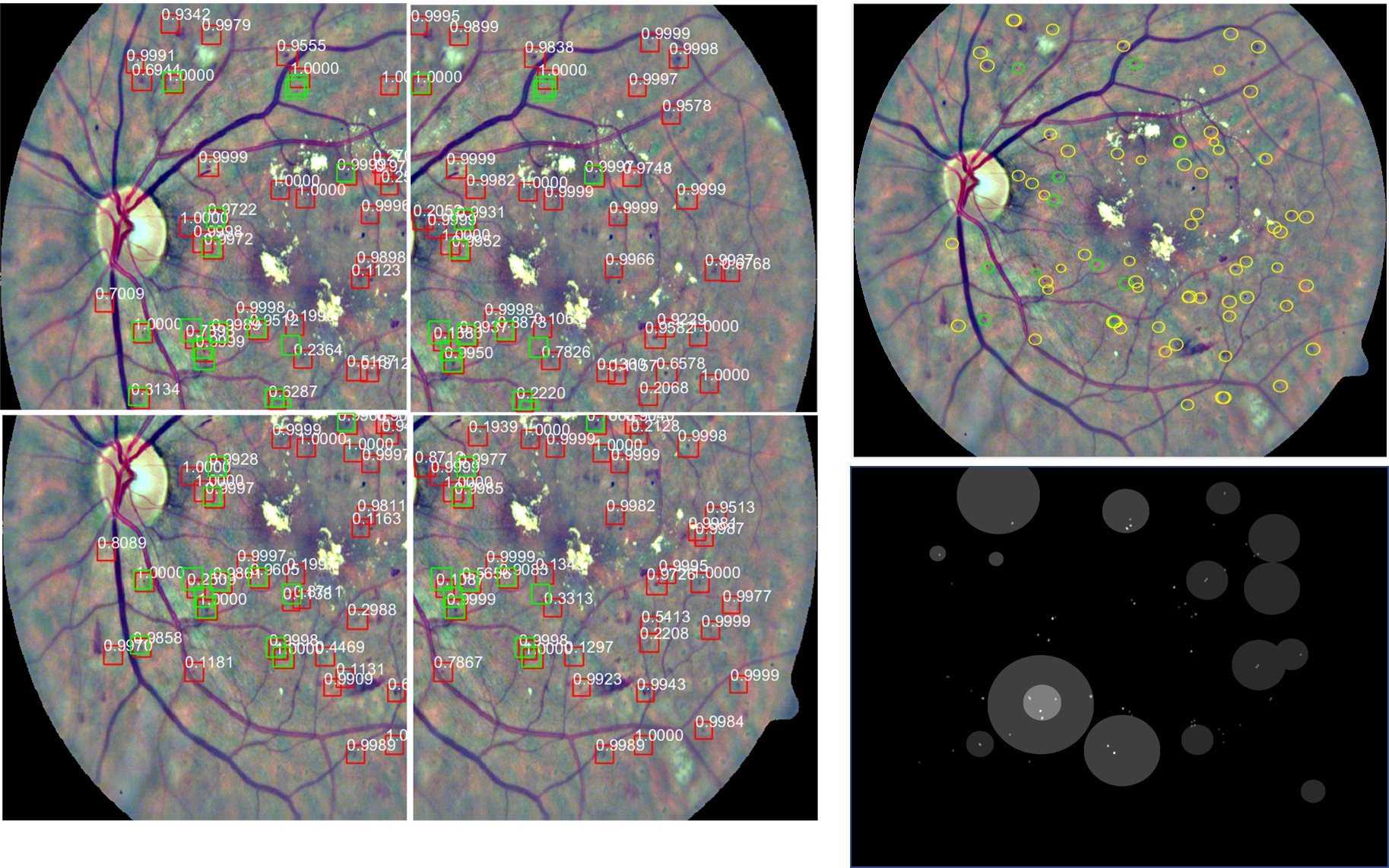}
	\caption{MAs detection results on the overlapped patches of image015 from the DiaretDB1 test set. At $\theta_{MA} \geq 0.1, and NMS_{MA} \geq 0.3$. Left: green rectangles show the detected TP MAs. Right(up): green circles show TP, yellow circles show FP and red ones show FN with confidence degree $\ge 75\%$. Right(bottom): ground truth with confidence degrees.} 
		\label{fig:results_2_b}
\end{figure}

\begin{figure}
	
		\centering
\includegraphics[width=1 \linewidth]{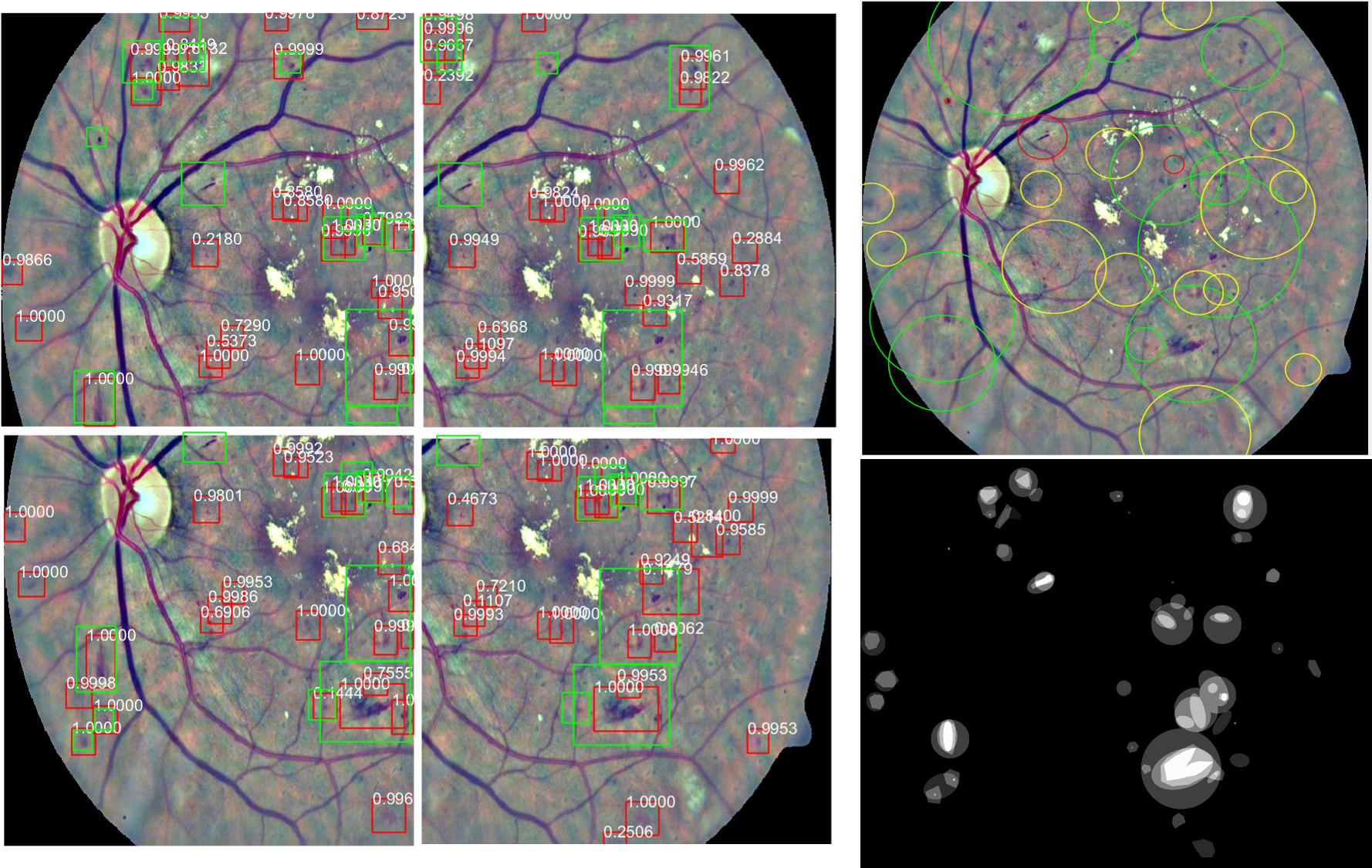}
\caption{HMs detection results on the overlapped patches of image015 from the DiaretDB1 test set. At $\theta_{HM} \geq 0.1 and  NMS_{HM} \geq 0.5$. Left: green rectangles show the detected TP HMs. Right(up): green circles show TP, yellow circles show FP and red ones show FN with confidence degree $\ge 75\%$. Right(bottom): ground truth with confidence degrees.} 
\label{fig:results_2_c}

\end{figure}

  For the private dataset (DCKAU), the method reached overall $FROC score_{{MA, HM}_{\theta=.1}} = 0.2995$ with $SN = 0.8956$ and $SP = 0.8173$, for moderate DR images and $FROC score_{{MA, HM}_{\theta=.1}} = 0.2641$ with $SN = 0.8699 and SP = 0.8091$ for severe DR images. These results are low compared to benchmark datasets such as e-ophtha due to low-resolution images where fovea detected as HM and also due to incorrect annotation as the expert stated that the method highlighted unannotated lesions. Figures \ref{fig:results_2_a}, \ref{fig:results_2_b}, and \ref{fig:results_2_c} show the results of per-lesions detection on two images: image015 taken from the DiaretDB1 dataset and C0003164 taken from the e-ophtha dataset. The method produces one single image, where both streams are used and the outputs of the branches are merged into one image at the last step. In Figure \ref{fig:results_2_a}, only MAs branch is used since the dataset has only MAs ground truth annotations.

\begin{table*}[ht]
	\centering
	\caption{FROC scores and sensitivity at different FPI of different methods on e-ophtha dataset.}
	\begin{tabular}{|l|l|c|c|c|c|c|c|c|c|}
		\hline
		\multicolumn{1}{|c|}{\multirow{2}{*}{Authors}} & \multicolumn{1}{c|}{\multirow{2}{*}{Methods}} & \multicolumn{7}{c|}{Sensitivity under different FPI values}                                                                & \multirow{2}{*}{FROC score} \\ \cline{3-9}
		\multicolumn{1}{|c|}{}                         & \multicolumn{1}{c|}{}                         & 1/8             & 1/4             & 1/2             & 1               & 2               & 4               & 8              &                      \\ \hline
		\shortstack{Orlando \\\cite{orlando2018ensemble}}                                     & CNN+ HEF+RF                                   & 0.1470          & 0.2030          & 0.2683          & 0.3680          & 0.4478          & 0.5187          & 0.6252         & 0.3683               \\ \hline
			\shortstack{Chudzik\\ \cite{chudzik2018microaneurysm1}}     &  FCN                             & \textbf{0.185}         & \textbf{0.313} &\textbf{0.465} & \textbf{0.604}&\textbf{0.716} &  \textbf{0.80}& \textbf{0.849}& \textbf{0.562} \\ \hline
		Wu \cite{wu2017automatic}                                            & Region growth+KNN                             & 0.063           & 0.117           & 0.172           & 0.245           & 0.323           & 0.417           & 0.573          & 0.273                \\ \hline
		Eftekhari \cite{eftekhari2019microaneurysm}                                      & Two- stages CNN & 0.091 & 0.258 & 0.401  & 0.534           & 0.579           & 0.667           & 0.771 & 0.471                \\ \hline
		Long \cite{long2020microaneurysms}                                         & \shortstack{Directional local\\contrast+\\Naive Bayesian}
		& 0.075           & 0.154           & 0.267           & 0.358           & 0.472           & 0.594           & 0.699          & 0.374                \\ \hline
		
			Chudzik \cite{chudzik2018microaneurysm2}     &  \shortstack{FCN+freezing \\interleaved layers}   &0.151& 0.264& 0.376& 0.468& 0.542& 0.595& 0.621& 0.431                     \\ \hline

			\shortstack{Proposed\\$\theta=.5$}& - &0.1247& 0.2273&0.2604  & 0.3953  & {0.5674 } & 0.7439 &0.7447 &\shortstack{FROC=0.4377\\AUC=0.8332 \\SN=0.7350  \\\textbf{SP=0.5650 }} \\ \hline
			
			\shortstack{Proposed\\$\theta=.3$}& - &0.1240 & 0.2234  &0.2605 & 0.3963 &0.5919 & 0.7462  & 0.7462  & 		\shortstack{FROC=0.4412\\AUC=0.82894  \\SN=0.7606   \\SP=0.5592} \\ \hline 
			
			\shortstack{Proposed\\$\theta=.1$}& - & {0.1217 }&  {0.2294 }  & {0.2681 }   &  {0.4211}   & {0.6094}  &  {0.7632 } & {0.7632} & \shortstack{  {FROC=0.4537}  \\\textbf{AUC=0.82897 } \\\textbf{SN=0.8157 }  \\SP=0.5514} \\ \hline

	\end{tabular}
\label{cpmema}
\end{table*}


\begin{table*}[ht]
	\caption{FROC score and sensitivity at different FPI of different methods on DiaretDB1-MA dataset.}
	\begin{tabular}{|l|l|c|c|c|c|c|c|c|c|}
		\hline
		\multicolumn{1}{|c|}{\multirow{2}{*}{Authors}} & \multicolumn{1}{c|}{\multirow{2}{*}{Methods}} & \multicolumn{7}{c|}{Sensitivity under different FPI values}                                                                                                                                                     & \multirow{2}{*}{FROC score}        \\ \cline{3-9}
		\multicolumn{1}{|c|}{}                         & \multicolumn{1}{c|}{}                         & 1/8                         & 1/4                         & 1/2                         & 1                           & 2                           & 4                           & 8                           &                             \\ \hline
			\shortstack{Orlando\\ \cite{orlando2018ensemble}}                                       & CNN+ HEF+RF                                   & 0.0516                      & 0.0877                      & 0.1198                      & 0.2885                      & 0.4450                      & \textbf{0.6142}                      & 0.7042                    & \shortstack{FROC=0.3301\\AUC = 0.9031} \\ \hline
		Adal \cite{adal2014automated}                                           & Semi-supervised learning                      & 0.024                       & 0.033                       & 0.045                       & 0.103                       & 0.204                       & 0.305                       & 0.571                       & 0.184                       \\ \hline
		Long \cite{long2020microaneurysms}                                           & \shortstack{Directional local contrast+\\ Naive Bayesian}   & 0.013                       & 0.026                       & 0.052                       & 0.104                       & 0.209                       & 0.400                       & 0.669                       & 0.210                       \\ \hline
			\shortstack{Chudzik \cite{chudzik2018microaneurysm1}}                                   & FCN                                           & \multicolumn{1}{l|}{\textbf{0.187}}  & \multicolumn{1}{l|}{\textbf{0.246}}  & \multicolumn{1}{l|}{\textbf{0.288}}  & \multicolumn{1}{l|}{\textbf{0.365}}  & \multicolumn{1}{l|}{\textbf{0.449}}  & \multicolumn{1}{l|}{\textbf{0.570} } & \multicolumn{1}{l|}{0.641}  & \multicolumn{1}{l|}{\textbf{0.392}}  \\ \hline
		\shortstack{Seoud \cite{seoud2016red}} & Dynamic shape features+RF                                         & \multicolumn{1}{l|}{N/A}  & \multicolumn{1}{l|}{N/A}  & \multicolumn{1}{l|}{N/A}  & \multicolumn{1}{l|}{N/A}  & \multicolumn{1}{l|}{N/A}  & \multicolumn{1}{l|}{N/A}  & \multicolumn{1}{l|}{N/A}  & \multicolumn{1}{l|}{0.3540}  \\ \hline
	
		\shortstack{Proposed\\$\theta=.5$}& - &0.0724   & 0.1029 &0.1304  & 0.1449  &0.2319 & {0.4493 } & 0.7464 & 		\shortstack{FROC=0.2683 \\AUC=0.74099 \\SN=0.8169 \\\textbf{SP=0.5189 }}  \\ \hline
		
		\shortstack{Proposed\\$\theta=.3$}& - &0.0656  & 0.0963 &0.1387  & 0.1533  &0.2482 & {0.4745 } & 0.7518 &\shortstack{FROC=0.2755 \\AUC=0.74099 \\SN=0.8344 \\SP=0.5172}  \\ \hline
		
		\shortstack{Proposed\\$\theta=.1$}& - &0.0709  & 0.0936 &0.1348  & 0.1489  &0.2553 & {0.4752 } & \textbf{0.7518 }& 		\shortstack{FROC=0.2758 \\\textbf{AUC=0.74099 }\\\textbf{SN=0.8589 } \\SP=0.5142 }  \\ \hline
	\end{tabular}
\label{cpmdma}
\end{table*}


\begin{table*}[ht]
	\caption{FROC score and sensitivity at different FPI of different methods on ROCh training dataset.}
	\begin{tabular}{|l|l|c|c|c|c|c|c|c|c|}
		\hline
		\multicolumn{1}{|c|}{\multirow{2}{*}{Authors}} & \multicolumn{1}{c|}{\multirow{2}{*}{Methods}} & \multicolumn{7}{c|}{Sensitivity under different FPI values}                                                                & \multirow{2}{*}{FROC score} \\ \cline{3-9}
		\multicolumn{1}{|c|}{}                         & \multicolumn{1}{c|}{}                         & 1/8             & 1/4             & 1/2             & 1               & 2               & 4               & 8              &                      \\ \hline
		\shortstack{Chudzik\\ \cite{chudzik2018microaneurysm1}}                                       &             FCN                                  & 0.039         & 0.067          & 0.141         & 0.174          & 0.243          & 0.306          & 0.385         & 0.193               \\ \hline

	\shortstack{Chudzik\\ \cite{chudzik2018microaneurysm2}}                                     & \shortstack{FCN+freezing \\interleaved layers}
			& 0.142& 0.201& 0.250& 0.325& 0.365& 0.390& 0.409 &0.298                \\ \hline
			
	\shortstack{Zhang\\ \cite{zhang2010detection}}      & \shortstack{Multi-scale\\ correlation of Gaussian}
	                    	& N/A& N/A& N/A& N/A& 0.126& 0.210& 0.243& 0.193               \\ \hline
	\shortstack{Javidi\\ \cite{javidi2017vessel}}& Discriminative dictionary
&	0.130& 0.147& 0.209& 0.287& 0.319& 0.353& 0.383& 0.261\\ \hline	
		
		\shortstack{Proposed\\$\theta =.5$} & -   &  {0.0682 } &  {0.1418 } &  {0.1561 } &  {0.2781} &  {0.4049} &  {0.6342}   &  {0.6781} & \begin{tabular}[c]{@{}l@{}}{FROC=0.3373}\\AUC=0.7352\\SN=0.6094\\ \textbf{SP=0.6111}\end{tabular}\\ \hline
		
			\shortstack{Proposed\\$\theta =.3$} & -   &   {0.0716} &  {0.1438} &  {0.1579} &  {0.2823} &  {0.4067} &  {0.6268}   &  {0.6938} & \begin{tabular}[c]{@{}l@{}}{FROC=0.3404}\\AUC=0.78968\\{SN=0.6238}\\SP=0.5988\end{tabular} \\ \hline	
 		
 			\shortstack{Proposed\\$\theta =.1$} & -   &  \textbf{0.0710} & \textbf{0.1407} & \textbf{0.1596} & \textbf{0.2817} & \textbf{0.4085} & \textbf{0.6385}   & \textbf{0.7230} &\begin{tabular}[c]{@{}l@{}}\textbf{FROC=0.3461}\\ \textbf{AUC=0.7877}\\ \textbf{SN=0.6595}\\ SP=0.5801\end{tabular} \\ \hline	
	\end{tabular}
\label{cpmdmaroch}
\end{table*}


\begin{table*}[ht]
	\caption{FROC score and sensitivity at different FPI of different methods on DiaretDB1-HM dataset.}
	\begin{tabular}{|l|l|c|c|c|c|c|c|c|c|}
		\hline
		\multicolumn{1}{|c|}{\multirow{2}{*}{Authors}} & \multicolumn{1}{c|}{\multirow{2}{*}{Methods}} & \multicolumn{7}{c|}{Sensitivity under different FPI values}                                                                & \multirow{2}{*}{FROC score} \\ \cline{3-9}
		\multicolumn{1}{|c|}{}                         & \multicolumn{1}{c|}{}                         & 1/8             & 1/4             & 1/2             & 1               & 2               & 4               & 8              &                      \\ \hline
		Orlando \cite{orlando2018ensemble}                                       &              HEF+RF                                   & \textbf{0.1930}& \textbf{0.3162} & \textbf{0.3807}& \textbf{0.4715} & 0.6425 & 0.7282 &  {0.7989}& \textbf{0.5044} \\ \hline
		
			\shortstack{Proposed\\$\theta=.5$}& - &0.0748  & 0.1423 &0.1806 & 0.3639  &0.6528 & {0.7889 } & 0.7889& 		\shortstack{FROC=0.4275 \\AUC=0.9487 \\SN=0.7362  \\\textbf{SP=0.9633}}  \\ \hline

			\shortstack{Proposed\\$\theta=.3$}& - &0.0726  & 0.1371 &0.1727  & 0.3565  &0.6546 & {0.7911 } & 0.7911 & 		\shortstack{FROC=0.4251 \\AUC=0.9487 \\SN=0.7422  \\SP=0.9592}  \\ \hline
			
			\shortstack{Proposed\\$\theta=.1$}& - &0.0748   & 0.1461 &0.1791   & 0.3719   &\textbf{0.6667}  & \textbf{0.8044} & \textbf{0.8044}&\shortstack{FROC=0.4353 \\AUC=0.9487 \\\textbf{SN=0.7552}  \\SP=0.9541 }  \\ \hline
	\end{tabular}
\label{cpmdhm}
\end{table*}



\newpage

\subsection{Image Level Detection Results}
Our method achieved $AUC_{MA=.5}=0.8332$ compared to $AUC_{MA=.1}=0.82897$ obtained by e-ophtha dataset. This is because at low $\theta$ more image are likely to be reported as abnormal due to FPs. For DiaretDB1-MA, the best $AUC_{MA=.1,.3,.5} = 0.74099$ compared to AUC = 0.9031 reported by \citet{orlando2018ensemble}. This is due to HMs annotation polygons used instead of exact annotation similar to annotations shown in Figure \ref{fig:screenshot012_hm} where the proposed method trained on exact lesions location and also because of some of small circular red lesions annotated as HMs which affect the overall performance either for MAs and HMs. For DiaretDB1-HM, $AUC_{HM{\theta=.1,.3,.5}}=0.9487$, this is due to the clear and obvious appearance of HMs unlike MAs, which make the chance of missing HMs low. Unlike other datasets, on ROCh dataset, the reported $AUC_{\theta=.1}=0.7877$, which is higher than $AUC_{\theta=.5}=0.7352$. This is because most of TP MAs appear at low $\theta$ due to of low images resolution.



\section{Discussion}

Compared to other methods, the proposed method generates fewer candidates,  as shown in Table \ref{train}. Also it performs better with early signs images that have only MA as shown in Tables \ref{cpmema} and \ref{cpmdmaroch} comparing with other methods because there is no pre-stage such as vessels segmentation nor high number of candidates.

In training, a critical issue of ground truth annotation of DiaretDB1 produced mistakes. Unlike other datasets, DiaretDB1 annotated lesions using geometrical shapes such as polygon and circles. This leads to incorrect lesion locations in patches due to annotation of pixels belong to the polygon as lesions pixels Hence, we eliminate patches that have empty ground truth. 

We noticed that some artifacts highlighted as lesions by experts lead to mistakes. The training and testing examples in e-ophtha are $C0001885, C0007104, C0007106, C0007156$ and in DiaretDB1 is  $image008$.

The proposed method performs better for MAs detection. The main reason is that it extracts candidates by grouping them into small and large lesions candidates but not MAs and HMs candidates, which leads to missing small HMs in detection. FCN-8 is used to extract medium to large HMs candidates; it fails to detect all MAs candidates due to their small size, leading to segmenting them with vessels pixels, especially those adjoining the vessels. Moreover, thin flame HM might not be detected accurately due to similarity with segmented blood vessels. We employ two different streams to overcome these issues: one for small red lesions and one for large red lesions.

Most false positive samples are related to laser treatment scars, lens dirt, speckles, and ends of thin blood vessels that are not segmented due to similarity with blot HM. Also fovea sometimes detected as HMs specially with low $\theta$. In DiaretDB1 and e-ophtha, the speckles and spots on the lens are not always distinguishable, but they are known by their fixed position on images. Hence, the most effective way to avoid them is to repair the camera lens. Moreover, it is hard to distinguish between artifacts and small lesions. Also, the quality of image affects detection, as has been noticed in the private dataset.

In general, when experts provide a lesion’s delineation and pixel annotation, several techniques are assessed on a per lesion basis. That means these approaches should detect every single lesion and analyze performance accordingly \cite{seoud2016red}. The performance of lesion-level detection should be high because the number of lesions and their locations and types are crucial to assess DR severity levels \cite{seoud2016red}. On the other hand, when image-level diagnosis only is offered, the diagnosis is per image instead of lesion pixels \cite{seoud2016red}. The per-image assessment is more interesting from the screening point of view because it evaluates the method’s performance in distinguishing images with DR signs \cite{seoud2016red}. Hence, lacking pixel-level annotation of datasets, such as MESSIDOR and IDRiD,  limits the testing assessment of the proposed system.

\newpage  
 \section{Conclusion}
 Retinal diseases are the most common cause of vision complications in adults and cause some symptoms, such as blurry vision. Sometimes, these symptoms can be very serious such as sudden blindness. In this paper, we employed deep learning techniques for red lesions detection and localization. We introduced a new technique based on deep learning for extracting candidates for large red-lesions. We employed two candidates generation methods, one for small red lesions and one for large red lesions, and two streams structure based on candidates type. For each stream, we used the VGG-16 model with ROI pooling,  regression, and classification layers. We evaluated the proposed method on different datasets for two different detection scenarios: lesion-level and image-level; the results show that it outperforms the state-of-the-art methods. We observed that what distinguishes the appearance of a lesion or image as abnormal is sometimes a complex set of interrelated elements on different scales; it is essential to exploit this information for better detection results. We plan to enhance vessels segmentation by using more datasets such as CHASE \cite{chase} Also, we are planning to grade images without pixel-level annotation, such as in MESSIDOR dataset, and extend our work to other lesion types such as exudates. Also we propose to extract fovea before HMs detection to decrease the number of false positives.

\paragraph{Conflict of interest}
The author has no competing interests to declare.

\bibliographystyle{plainnat}
\bibliography{ref_}

\end{document}